\documentclass[12pt]{iopart}
\usepackage{graphicx}
\usepackage{lineno}
\usepackage{iopams}
\begin{document}
\title[Multiscale periodicities in aerosol optical depth over India]
{Multiscale periodicities in aerosol optical depth over India}

\author{S. Ramachandran$^1$, Sayantan Ghosh$^{2,3,}$$^{\dagger}$, Amit Verma$^4$ and P. K. Panigrahi$^3$}

\address{$^1$ Physical Research Laboratory, Navrangpura, Ahmedabad 380009, India\\ 

$^2$ School of Chemistry and Physics, University of KwaZulu-Natal, Durban 4000, South Africa\\ 

$^3$ Dept. of Physical Sciences, Indian Institute of Science Education and Research Kolkata, 
Mohanpur 741252, India \\

$^{\dagger}$ Currently at SUPA, School of Physics and Astronomy, University of St Andrews, North Haugh, St Andrews KY16 9SS, Fife, UK\\ 

$^4$ Department of Computer and Information Science and Engineering, University of Florida, Gainesville, Florida 32611, USA} 

\ead{ram@prl.res.in}

\begin{abstract}

Aerosols exhibit periodic or cyclic variations depending on 
natural and anthropogenic 
sources over a region, which can get modulated by synoptic meteorological parameters  
such as winds, rainfall and relative humidity, and long-range transport.
Information on periodicity and phase in aerosol properties
assumes significance in prediction as well as to examine the
radiative and climate effects of aerosols including its association with
changes in cloud properties and rainfall.
Periodicity in aerosol optical depth, which is a columnar measure of 
aerosol distribution, is determined using continuous
wavelet transform over 35 locations (capitals of
states and union territories) in India. 
Continous wavelet transform is used in the study because it is better suited to extract the periodic 
and local modulations present at various frequency
ranges, as these features are invisible in 
conventional methods such as Fourier Transform.  
Monthly mean aerosol optical depths (AODs) 
from MODerate Resolution Imaging Spectroradiometer (MODIS) on board the Terra 
satellite at 1$^o$$\times$1$^o$ resolution from January 2001 to
December 2012 are used.
Annual and quasi-biennial oscillations (QBO) in AOD are evident 
in addition to the weak semi-annual (5-6 months) and 
quasi-triennial oscillations ($\sim$40 months). 
The semi-annual and annual oscillations
are consistent with the seasonal and yearly cycle of
variations in AODs. QBO type periodicity in AOD
is found to be non-stationary while the annual period is stationary. 
The 40-month periodicity indicates the
presence of long term correlations in AOD.
The observed periodicities in MODIS Terra AODs are 
also evident in the ground-based AOD measurements made over Kanpur 
in the Indo-Gangetic Plain.
The phase of the periodicity in AOD is stable in the mid-frequency range, 
while local disturbances in the high-frequency range and 
long term changes in the atmospheric composition 
give rise to unstable phases in low-frequency range. 
That modulations in AOD over one location/region can influence 
the other is revealed by the presence of phase relation among 
different locations.
\end{abstract}


\section{Introduction}

Atmospheric aerosols exert a cooling effect on the Earth's climate 
through direct and indirect effects which partially offset the 
warming caused due to greenhouse gases.  
The sources of aerosols can be natural (dust, sea salt, biogenic and volcanic) and 
anthropogenic (combustion of fossil fuel from urban/industrial processes and biomass burning).
Dust, sea salt and sulfate produced over the ocean surfaces dominate 
the natural global aerosol abundance, however, a fraction of dust 
in the atmosphere could be due to anthropogenic activities (Habib et al., 2006; Prospero et al., 2002). Similarly, 
smoke from natural burning such as due to forest fires  
is treated as natural component of biomass burning; while the 
burning of fuel wood, dung cake and crop waste burning 
are anthropogenic processes.  
Atmospheric aerosols
modify the Earth-atmosphere radiation budget by scattering
and absorbing the incoming solar radiation (direct effect), and
the processes of formation of clouds and precipitation (indirect effect).
The direct and indirect aerosol radiative effects remain a 
significant uncertainty in climate studies (Solomon et al 2007).  

Scattering (sulfate) and absorbing (black carbon) particles 
cool the Earth's surface, 
however, their radiative effects in the atmosphere vary with altitude. 
For scattering particles, 
the top of the atmosphere forcing is almost the same as the surface forcing; 
while for absorbing aerosol species the surface forcing is about 2 to 3 times larger than the top 
of the atmosphere forcing,  
which gives rise to a large atmospheric warming. 
The greenhouse gases are longer lived, globally well mixed and their radiative 
effects are homogeneous and 
one of warming throughout the atmosphere starting from the surface. 
In contrast, aerosols reside in the atmosphere for about a week, exhibit regional signatures 
and can either warm or cool the atmosphere. 
Although aerosols are abundant near source regions, 
they impact global climate as aerosols and their radiative influence can be  
transported to other regions due to atmospheric circulation. 
On temporal scales, the forcing due to aerosols is greatest during daytime 
and in summer. In contrast, the greenhouse gas forcing acts over 
the full diurnal and seasonal cycles. Thus, 
aerosols perturb the Earth-atmosphere radiation budget 
differently than greenhouse gases. 

The most important characteristics required to estimate the 
radiative influence of aerosols are 
aerosol optical depth (AOD), single scattering albedo (SSA) 
and asymmetry parameter.
Aerosol optical depth is the most crucial parameter 
to study aerosol-climate interaction among the three, 
because 
aerosol radiative forcing changes due to 
increase in AODs overwhelm the forcing changes 
due to the increases in single scattering albedo and the asymmetry parameter values 
(Ramachandran 2005).
SSA (ratio of scattering to extinction) values 
can range from 0 (absorber) to 1 (scatterer). 
The aerosol radiative forcing at the 
surface 
is nearly linearly related to AODs. For the same AOD value when SSA is 
lower, the surface forcing is higher and the atmospheric forcing also 
becomes larger.   
In addition, the radiative forcing at 
the top of the atmosphere due to lower SSA 
can become positive when surface albedo is higher (Solomon et al 2007). 
Thus, it is clear that though aerosols with lower SSA have the potential 
to change sign of the forcing at the top of the atmosphere, AOD is more crucial 
in aerosol-climate impact investigations owing to its linear dependence to radiative forcing.    

Aerosol optical depth is dependent on a number of factors
including the aerosol burden throughout the atmospheric column, the aerosol size
distribution, and the chemical composition (as it relates to water uptake and refractive index).
Aerosol optical depth from the Moderate Resolution Imaging Spectroradiometer (MODIS) 
satellite is found to exhibit seasonal variations.  
Biomass burning aerosol is significant over the Gulf of Guinea region during 
January-March, which shifts to southern Africa during August-October (Solomon et al 2007). 
Transport of mineral dust from Africa to south America occurs during 
January-March while mineral dust gets transported over west Indies and central America 
during August-October. 
AODs are found to show a minimum in winter (January) and a maximum during 
summer (April-July) (Jin et al 2005; Ramachandran and Cherian 2008). 
This indicates that AODs can exhibit periodic or cyclic variations depending on natural and 
anthropogenic   
sources over a region, which can get modulated by the synoptic meteorological parameters  
such as winds, rainfall and relative humidity. 
Transport of dust and sea salt
from the adjacent source regions during different seasons can also modulate the 
periodic nature of AODs (Prospero et al 2002; Kaufman et al 2002). 
The knowledge on the periodicity of aerosol characteristics 
will be useful in statistical forecasting of visibility and air quality. 
In addition, information on the periodical nature of aerosol properties along with the 
knowledge on the causes for their periodicity, can be 
useful in the prediction of climate impact due to aerosols. 

Wavelet transform has emerged as a powerful tool to study the 
transient and time varying phenomena (Daubechies 1992; Mallat 1999). 
It is an ideal tool for identifying variations at multiple scales. In particular, 
continuous Morlet wavelet has optimal sensitivity making it a good choice for 
observation of long term periodicities. The optimal sensitivity of wavelets to 
both high and low scale variations has already found application in the 
merger of different proxy data for establishing weather variations on 
the millennial scale (Moberg et al 2005). 
In this study for the first time,   
the periodicities in AODs and their phase over a large spatial domain (India) 
governed by different aerosol sources and long-range transport are derived. 
Wavelet transforms are used 
to probe both the high- and low-frequency periodic components in AOD 
over different regions of India and their phase relationships.

\section{Study region}

Asia accounts for about 60\% of the world's population, and faces 
serious environmental threats in terms of air pollution, monsoon floods, droughts and 
associated climate change.
Increases in aerosol loading due to 
growing population and industrialization in recent decades 
have resulted in an increase in health-related problems, 
and impacted air quality, agriculture and water resources in Asia (Lau et al 2008). 
Anthropogenic aerosols over south and east Asia can significantly 
change the energy balance of the Earth-atmosphere system 
(Lau et al 2008; Solomon et al 2007). 
Accumulation of absorbing aerosols (dust and black carbon) during the 
pre-monsoon season over the Tibetan Plateau can influence the 
Asian and Indian summer monsoon rainfall (Lau and Kim 2006).  
India is densely populated (population $>$1 billion), industrialized 
and in recent years has witnessed
impressive economic development. 
On an annual mean scale 
7-year (2000-2006) analysis of MODIS AODs showed that 
on annual mean scale AODs over India were the highest (Remer et al 2008, Table 2) 
when compared to the rest of the 
world. 
With increase in urbanization
the usage pattern of fossil and bio fuels can change leading to changes
in aerosol properties, which may influence precipitation and
can spin down the hydrological cycle (Lau et al 2008; Solomon et al 2007). 

Bio fuels such as fuel wood, dung cake and crop waste are used in rural areas, the
emissions of which predominantly contribute to aerosol formation (Habib et al 2006).
Aerosol emissions from fossil fuels such as coal, petrol and diesel oil
dominate the urban areas. These fossil fuels are used
in electrical power generation, iron and steel production,
oil refining and petrochemical processes, domestic usages and
transportation.
The Indian subcontinent apart from being a source region for aerosols, is
bordered by densely populated and industrialized
areas on the east and western sides 
from where different aerosol species such as
mineral dust, black and organic carbon,
nitrates, sulfate particles and organics are produced and transported, and therefore
is one of the regional aerosol hot spots.
The Indian landmass comprises
coastal regions, inland plains, semi arid regions, mountains and plateau regions.
The Indian subcontinent
experiences tropical and subtropical climatic conditions resulting in
extreme temperatures, rainfall and relative humidity.
These features introduce large variabilities in aerosol characteristics on spatial
and temporal scales over India (Ramachandran and Cherian 2008; Habib et al 2006).
In this context, it is
important to determine the periodicities in aerosol characteristics 
over a regional aerosol hot spot which can be helpful
not only in regional scale, but also in global air quality and climate.

\section{Data}

The MODerate Resolution Imaging Spectroradiometer (MODIS) is a 
remote sensor on board the Earth Observing System
(EOS) Terra and Aqua satellites.
MODIS Terra and Aqua satellites operate at an altitude of 705 km
with the Terra spacecraft crossing the equator at about 1030 LST
(ascending northward) while the Aqua spacecraft crosses
the equator at around 1330 LST (descending southward) (King et al 2003; Remer et al 2008).
AOD data are available from Terra since March 2000, while AOD data 
are available from Aqua starting from July 2002. 
Level 3 MODIS Terra Collection 5.1 quality assured (QA)
monthly average 0.55 $\mu$m AOD
at 1$^{o}\times$1$^{o}$
data from
January 2001 to December 2012 are utilized in this study, 
to maintain uniformity and because more number of data points on temporal scale 
is available from Terra.
Validation and comparison of AODs retrieved from Terra and Aqua, with ground-based
AErosol RObotic NETwork (AERONET) sun/sky radiometer measured
AODs (Holben et al 2001) over the globe revealed that
AODs from Terra and Aqua showed only little
differences and
agree very well over ocean and land (Remer et al 2008).
MODIS level 3 atmospheric products are sorted into 1$^o$$\times$1$^o$ cells on an
equal-angle global grid from level 2 atmospheric products that span over 24h period
(King et al 2003).
MODIS retrieval algorithms attempt to match the MODIS observed
surface reflectances to a look-up table of precomputed
reflectances for a wide variety of commonly observed
aerosol conditions (King et al 1999) over land and ocean.
The
predicted retrieval uncertainty of MODIS derived AODs over land is
$\pm$ (0.05 + 0.15AOD)
(Remer et al 2008).

MODIS AOD data corresponding to the capitals of 28 states and 7 union territories in India
divided into seven regions are analyzed and discussed (Table 1, Ramachandran and Cherian, 2008).
The division is based on geography and meteorological conditions, and within the context of
urban and rural development patterns.
The capitals of the states and union territories are chosen because 
most of these locations are urban centers, 
have medium to dense population (based on 2001
Indian census). The capitals across different regions are 
governed by different aerosol sources. 
For example, Delhi and Mumbai in addition to being densely populated,  
they are also largest commercial centers in India. 
The metro cities (population $>$10 million) such as 
Delhi, Mumbai, Kolkata and Chennai are sources of urban/industrial 
and automobile emissions. 
In contrast, northeast India is sparsely populated and is 
rich in natural resources of oil and gas. 

Seasonal mean climatology of aerosol optical depth distribution over India 
calculated from the 12-year (2000-2012) MODIS Terra AOD data, and synoptic wind patterns 
are plotted in Figure 1. 
The mean
synoptic surface winds vary as a function of season (Figure 1). 
During winter (December-January-February)
the winds are calm, north/northeasterly and
are from the northern hemisphere. During the southwest summer monsoon
(June-July-August-September),
the winds are stronger, moist and come from the marine and western
regions surrounding India (Figure 1). The winds are in a transitory phase and start shifting in direction
during post-monsoon (October-November) from
southwest to northeast. During the pre-monsoon season (March-April-May) the winds
originate and travel from the west of Indian subcontinent.
AOD exhibits significant regional and seasonal variations 
across India (Figure 1). 
AODs are higher in monsoon when compared to post-monsoon. 
AODs are higher over the Indo-Gangetic Plain 
throughout the year when compared to the rest of India. 
AODs are higher over the Indo-Gangetic Plain during winter and post-monsoon  
because of the dominance of fine mode aerosols from fossil fuel and 
biomass burning, while the higher AODs during pre-monsoon and monsoon 
are attributed to the dominance of coarse mode dust and sea salt particles. 

Seasonal variations in AOD are found to 
be significant over northwest, north and east India (Ramachandran 
and Cherian 2008). Annual mean AOD over northeast India
was lowest (0.3) and the contribution of 
fine mode aerosols (aerosols of size $<$1 $\mu$m) to the 
optical depth was higher (0.95) (Ramachandran and Cherian 2008). 
Northeast India is meagerly populated, and aerosols from natural burning of 
forest fires and biomass burning dominate. Annual mean AOD 
over western India is higher while the fine mode fraction was lower ($\sim$0.6) 
(Ramachandran and Cherian, 2008). Over western India the fine mode 
contribution to the aerosol optical depth is lower as the region is 
influenced by both natural (transport of mineral dust and sea salt) 
and anthropogenic sources (fossil fuel and biomass burning combustion).    
An analysis on aerosol optical depth and its periodicities over locations 
governed by different aerosol sources is expected to 
provide information on the spatial and 
temporal variations of aerosol sources and their influence on AOD, 
which are key inputs while examining the 
radiative and climatic effects of aerosols.

\section{Methodology}

In this section details about continuous wavelet transform (CWT) and  
the phase information obtained from the complex continuous wavelet transform are outlined. 
Wavelet transform, since its advent has been a valuable tool for signal processing 
(Daubechies 1992; Torrence and Compo 1998). 
It has been applied to analyze signals in different fields of science   
including geophysics, for example, to study El Ni\~no Southern Oscillations (Torrence and Compo 1998), 
tropical convection over the western Pacific (Weng and Lau 1994) among others. 
Detailed description on the use of wavelet transforms in geophysics 
can be found in Foufoula-Georgiou and Kumar (1995). 
The CWT of a data set $X=\{x_i\}, i \in \mathbb{Z}^+$ is given by,

\begin{linenomath}
\begin{equation}
W_i(s)=\sum_{j=1}^{N-1} x_j \psi^*\left(\frac{i-j}{s}\right)
\label{eq:eq1}
\end{equation}
\end{linenomath}

where $s$ is the scale and $N$ is the data length. $\psi(s)$ is a well localized 
(in both physical and Fourier domains), zero mean and integrable function and 
is called the \textit{mother wavelet}. Eqn. 1 represents a convolution equation, 
where the wavelet coefficients are calculated by convolving the scaled and translated versions 
of $\psi(n)$ with $x_i$. Thus, it is clear that $s$ is the scaling parameter 
and $j$ the translation parameter. In this analysis, Morlet wavelet is utilized, whose real form is given by,

\begin{linenomath}
\begin{equation}
\psi(n)=C\cos(5n)e^{-\frac{n^2}{2}}
\end{equation}
\end{linenomath}

where C is a normalization constant. This function has a wide support 
and allows to get more accurate results from the computationally 
performed convolution. Since this function is real it can be used to  
retrieve periodic structures from the AOD data. In order to 
obtain the phase relationships between various stations  
the complex Morlet function is used. The complex Morlet function is expressed as 

\begin{linenomath}
\begin{equation}
\psi(n)=\frac{1}{\sqrt{\pi F_b}}\exp \left(2\imath \pi F_c n-\frac{n^2}{F_b}\right)
\end{equation}
\end{linenomath}

where $F_b=1$ and $F_c=1.5$ are the bandwidth parameter and wavelet 
center frequencies respectively (see Teolis 1998). The phase angle is given by

\begin{linenomath}
\begin{equation}
\phi(n)=\tan^{-1} \left(\frac{\mbox{Im}[\psi(n)]}{\mbox{Re}[\psi(n)]}\right)
\label{eq:eq2}
\end{equation}
\end{linenomath}

Since the data length is 12 years (144 months), the analysis is limited to a scale of 32. 
As shown in Figure 2, the cone of influence is big enough 
to make the wavelet coefficients significant and reliable at this scale. 
The results obtained using the real  
Morlet wavelet (Eqn. 2) are used to extract the multiscale periodicities in aerosol optical depths, 
while the complex Morlet function is used to establish the phase relations, if any, 
between the study locations. Periodicity in atmospheric/geophysical parameters can provide information 
on the cyclic/periodical behavior and local modulations of the parameter, while the phase relation 
can be used to understand the influence aerosols 
on the local scale exert on a regional scale and/or surrounding locations.      

\section{Results and Discussion}

Results pertaining to four locations in India for a comparative study are highlighted in the study. 
Mumbai and Chennai represent the western and eastern coastal regions with 
Bengaluru lying between them (Figure 1). These three locations are interesting because of high 
urbanisation and industrialization, which allows one to observe the anthropogenic influence in 
aerosol characteristics, in conjunction with their vastly different ambient 
environments (Mumbai and Chennai are urban cities located near the coast, while Bengaluru 
is a continental location) and climatic conditions (Figure 1). 
Kohima is chosen because it is less industrialized and 
its atmosphere is dominated by fine mode aerosols 
produced by natural biomass burning (forest fires). 
Kohima and Chennai AODs are lower than Bengaluru and Mumbai (Figure 3). 
Three 
dominant periods in the signal can be seen, one corresponding to 12 months and the 
others corresponding to approximately 24 and 40 months. 
A period of approximately 5-6 months is observed at all 
the locations, though it is not significant as the other periodicities.
This is indicative of an external influence (over the natural cycle) in AODs. 
The 5-6 and 12-month period is consistent 
with the seasonal (summer high, winter low) and 
annual patterns  
seen in AODs (Jin et al 2005; Ramachandran and Cherian 2008).
These dominant periods are observed in all the other study locations also (not shown).

Figure 3 shows the monthly 
averaged AOD data and the local variation in various geographical regions, 
obtained using Morlet wavelet technique for Chennai (south India), 
Mumbai (west India), Bengaluru (south India) and Kohima (northeast India). 
Though the 12 month period is a stationary period (i.e., it does not vary over time), 
the 24 month period is non-stationary. 
The 24-month non-stationary periodicity can be associated with  
quasi-biennial oscillation (QBO). 
Quasi-biennial oscillation refers to downward propagating easterly or
westerly winds in the equatorial stratosphere ($\sim$16-50 km). 
This non-stationary periodic 
nature in AOD is consistent with QBO variation because the periodicity of 
QBO itself varies from 22 to 34 months. 

QBO is primarily a stratospheric phenomenon 
and its influence on columnar AODs is expected to be less. However, it has been shown that 
columnar AODs can get influenced by the different phases of 
QBO 
(Beegum et al 2009) over low latitudes or around the equator. 
The meridional circulation induced by the QBO in winds 
accompanied by convection (subsidence) over the equator during the 
east (west) phase with a cooler and higher tropopause (warmer and a lower tropopause),  
and the associated vertical and horizontal 
mixing of mass flux can modulate the AODs (Beegum et al 2009). 
This suggests that QBO, despite primarily being a stratospheric feature, can 
play a role in influencing the constituents in the troposphere including aerosols. 
Aerosols in the upper troposphere-lower stratosphere altitude 
region (10-30 km) over tropical latitudes contribute on an average  
of $\sim$12\% to the columnar AOD (Kulkarni et al 2008). 
It has also been found that the contribution of aerosols in the 
10-30 km exhibit seasonal variations, and their contribution to the 
columnar AODs can vary between 10 and 20\% (Kulkarni et al 2008), 
consistent with the finding on increase/decrease in AODs during the  
different phases of QBO. 
A 40 month cycle in AOD is also seen, which  
appears only after sufficient averaging (Figure 3). This implies the presence of 
long term correlations in AOD over the study locations. 
The results from the present study on  
the QBO and 40-month periodicities in AODs agree with 
Beegum et al (2009). 

An analysis of the monthly mean 0.50 $\mu$m AODs during 2001-2012 
from Kanpur (26.5$^{o}$N, 80.2$^{o}$E, 123m
above mean sea level (AMSL)), measured using ground-based AERONET 
sun/sky radiometers (Holben et al 2001), is undertaken in order to 
examine whether the observed periodicities are evident 
in ground-based measurements of columnar AOD as well. 
Kanpur is an urban, industrial and densely populated city 
with
a population of more than 4 million and located $\sim$250 km east of the
mega city, New Delhi (Figure 1). Kanpur AODs are mostly in the range of 0.5 to 1  
(Figure 4). 
The analysis reveals that AODs obtained between 2001 and 2012 over Kanpur 
also exhibit $\sim$5-6, $\sim$12, $\sim$24 and $\sim$36 month 
periodicities (Figure 4) 
consistent with periodicities derived using remote sensing AODs over 
different locations in India.  

In Figure 5, the phase relations obtained using Eqn. 4 are plotted. 
It is interesting to see 
a stationarity in the phases at scale 20. 
In contrast, at scale 10, which is the high-frequency range, 
non-stationarity in the phases is observed. 
All the stations 
undergo a phase variation around the month 35 (at scale 30). 
The phase relations also provide 
a measure of the recovery time to their original phases  
at various locations; Chennai recovers the fastest while 
the other stations take more longer to recover. 
The low-frequency range is also non-stationary. 
This makes the analysis of their phase differences important. 
In Figure 6 the
phase differences between the various stations are depicted. 
At scale 20 (mid-frequency range) 
AODs in the study locations are periodically in and out of phase, 
while
Chennai stays in phase with Mumbai and Bengaluru most of the time. 
It should be noted that $\Delta \phi = 0$ means zero phase lag, 
while $\Delta \phi > 0$ means the first station leads the phase 
and $\Delta \phi < 0$ means the second station leads the phase. 
The observation about the recovery time in Figure 5 is 
corroborated by Figure 6. As the individual phases 
take a long time to recover, the phase differences also take  
long time to recover. This long recovery time could be due to an 
external modulation/forcing from the urban/industrial emissions in these 
locations, which is a constant source and whose intensity can exhibit variations.   

\section{Conclusions}

Multiscale periodicities and phase in aerosol optical depths are determined by performing 
an analysis  
over 35 locations (capitals of 
states and union territories) in India. Continuous wavelet transform 
is chosen 
to extract the periodic and local modulations present at various frequency 
ranges which are invisible in the conventional methods such as Fourier Transform. 
Monthly mean aerosol optical depths from MODIS on board the Terra  
satellite at 1$^o$$\times$1$^o$ resolution from January 2001 to 
December 2012 are used.

The major findings of the study are: 

The study reveals the presence of 
5-6 and 40 month periods in AOD, in addition 
to the annual and quasi-biennial oscillations over the study locations.
The semi-annual and annual oscillations  
are consistent with the seasonal and yearly cycle of 
variations in AODs. QBO type periodicity in AOD 
is non-stationary consistent with the variation in 
the temporal occurrence of QBO. The 40-month periodicity suggests the 
presence of long term correlations in AOD.   

AODs obtained from ground-based measurements 
over Kanpur, an urban, densely populated location in the Indo-Gangetic Plain, 
also showed semi-annual, annual, QBO and 40-month periodicities 
consistent with the periodicities observed in MODIS Terra AODs.

The phase of periodical behavior is stable in 
the mid-frequency range, while in the high- and low-frequency ranges 
the phases are unstable implying time-local disturbances in the high-frequency range.  
The instability in the low-frequency range could be attributed to long term changes in 
the atmospheric composition and structure over the locations due to various 
factors such as industrial emissions and changes in forest cover for example. 

The presence of phase relation 
among different locations signifies that 
the modulations in AOD over 
one location/region can affect the other. 

Results from this study on the 
quantitative determination of the periodic nature of aerosol 
characteristics over a large spatial domain governed 
by a variety of aerosol sources will be useful 
while conducting visibility and air quality studies, and in  
global climate simulations and/or predictions of 
aerosols and their radiative effects.   

This study should be extended to the other regions of the globe   
to ascertain and correlate the natural and 
anthropogenic sources/processes contributing to the cyclic 
behavior in aerosol characteristics. \\

\noindent{\bf Acknowledgments}

Monthly mean MODIS Terra aerosol optical depths are 
downloaded from the GES-DISC, NASA. 
Winds are downloaded from http://www.cdc.noaa.gov. 
Thanks are due to Dr. Sumita Kedia, CDAC, Pune 
for her help in drawing Figure 1.
We thank  B.N. Holben, R.P. Singh and S.N. Tripathi for their efforts in establishing and
maintaining the AERONET sun/sky radiometers at Kanpur,
from which monthly mean aerosol optical depth at 0.5 $\mu$m are used 
in the study. \\

\newpage
\begin{figure}[h]
\includegraphics[width=1.0\textwidth]{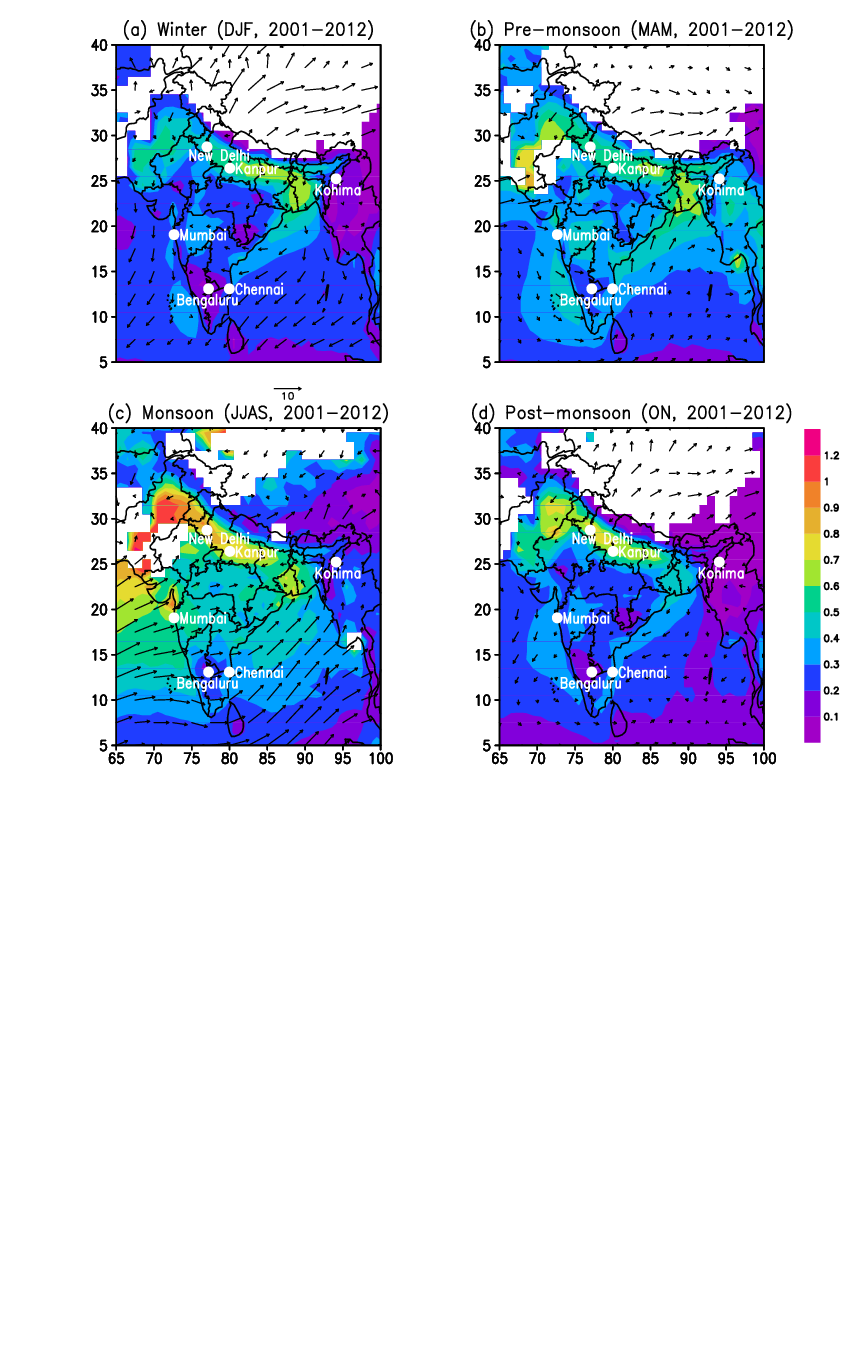}
\vspace*{-4in}
\noindent \caption{\it Seasonal mean climatology (2000-2012) of aerosol optical depths and surface winds
over India during (a) winter (December-January-February, DJF), pre-monsoon (March-April-May, MAM), 
(c) monsoon (June-July-August-September, JJAS) and (d) post-monsoon (October-November, ON).  
The shaded contours correspond to 0.55 $\mu$m aerosol optical depths, on which surface
winds (ms$^{-1}$) represented by arrows
are overlaid.}
\end{figure}

\newpage
\begin{figure}[h]
\includegraphics[width=1.0\textwidth]{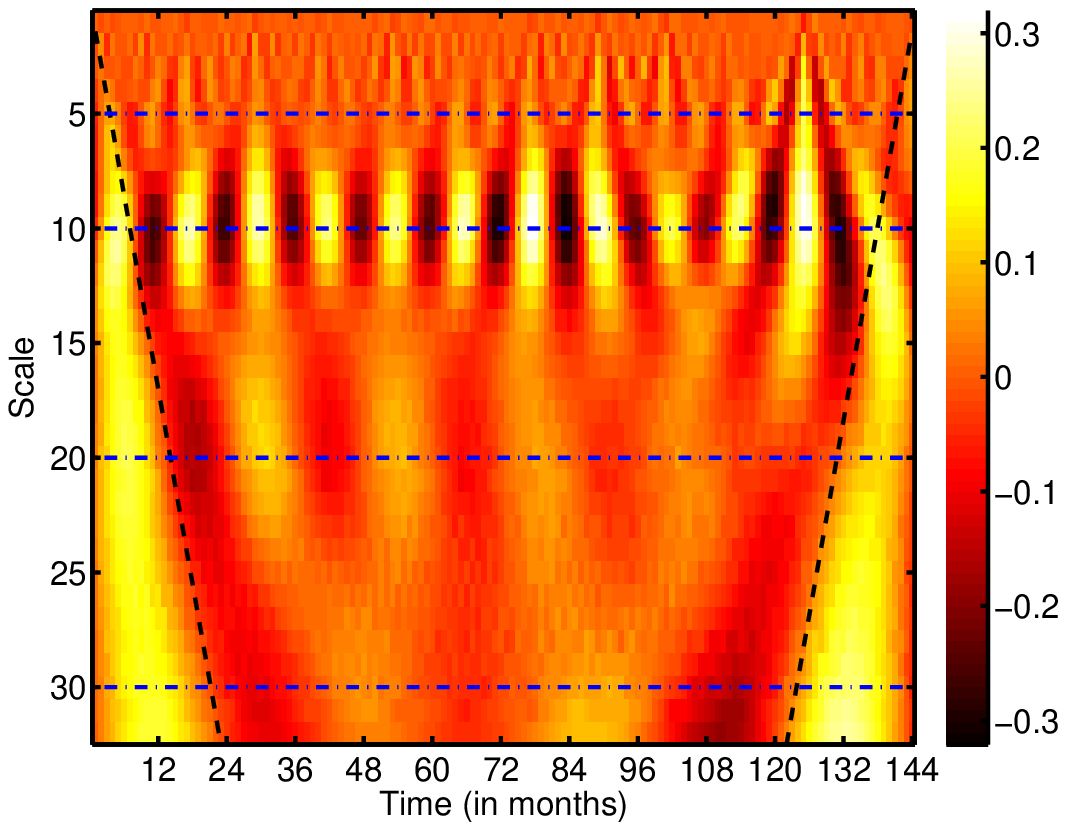}
\noindent \caption{\it The scalogram for Kohima upto scale 32. 
The cone of influence (black dotted line) is shown. The oscillations 
(drawn as blue dotted lines) at semi-annual (at scale 5), 
annual (at scale 10), QBO (at scale 20) and 
40 month (at scale 30) are clearly visible.}  
\end{figure}

\newpage
\begin{figure}[h]
\includegraphics[width=1.0\textwidth]{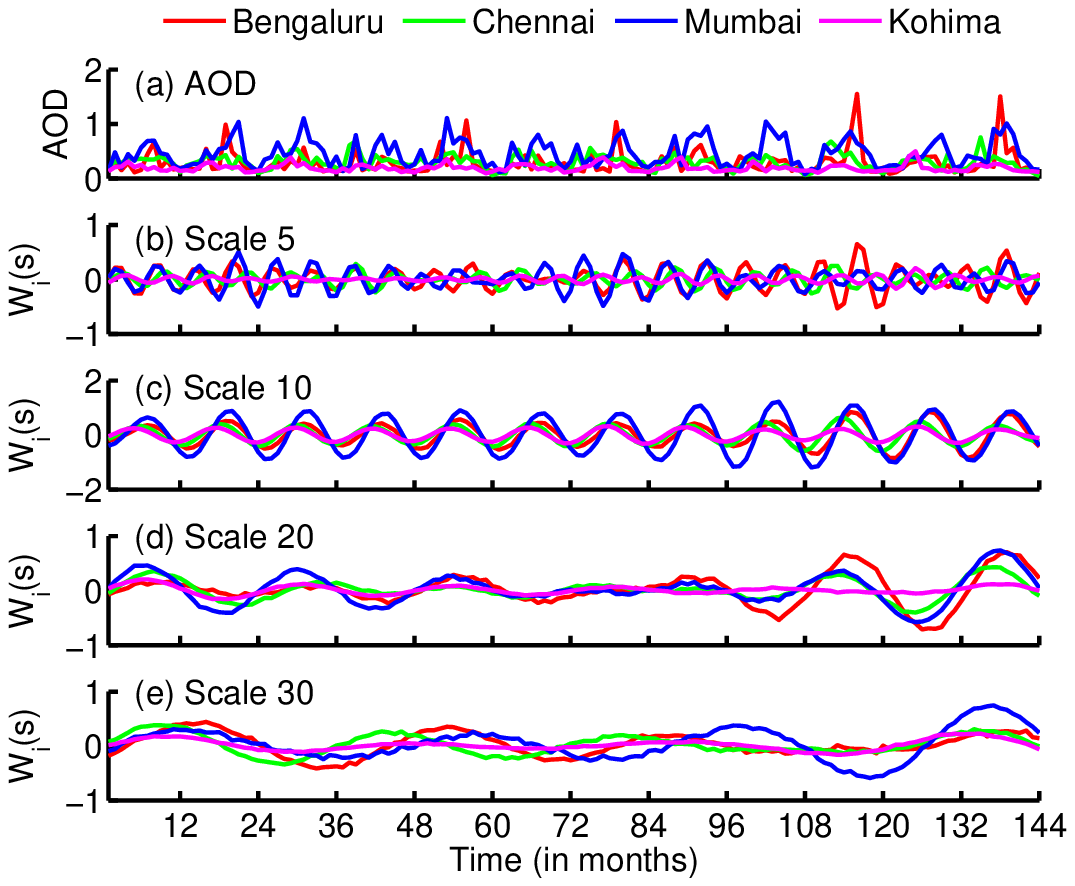}
\noindent \caption{\it (a) Monthly mean aerosol optical depths over Bengaluru, 
Chennai, Mumbai and Kohima from January 2001 to December 2012. Wavelet coefficients over Bengaluru,
Chennai, Mumbai and Kohima at (b) scale 5, (c) scale 10, (d) scale 20 and (e) scale 30.}
\end{figure}

\newpage
\begin{figure}[h]
\includegraphics[width=1.0\textwidth]{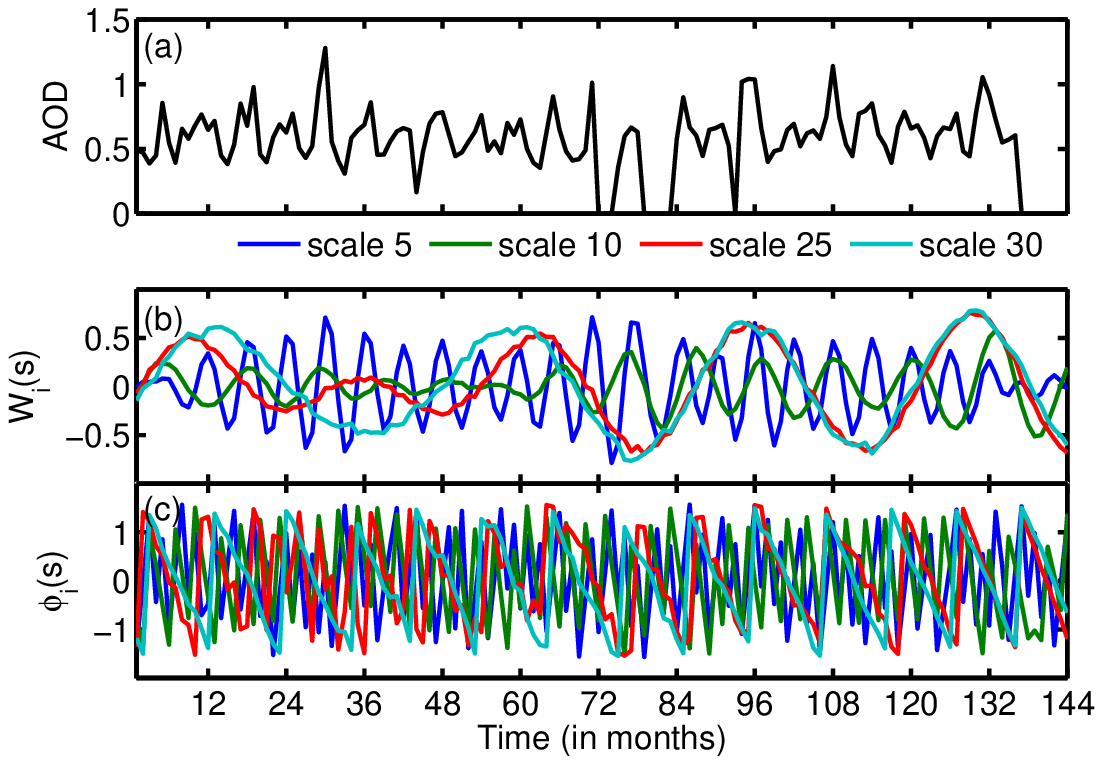}
\noindent \caption{\it (a) Monthly mean aerosol optical depths over Kanpur during 2001-2012. 
(b) Wavelet coefficients of aerosol 
optical depths obtained over Kanpur at different scales. 
(c) Phase relation of aerosol optical depths over Kanpur.}
\end{figure}

\newpage
\begin{figure}[h]
\includegraphics[width=1.0\textwidth]{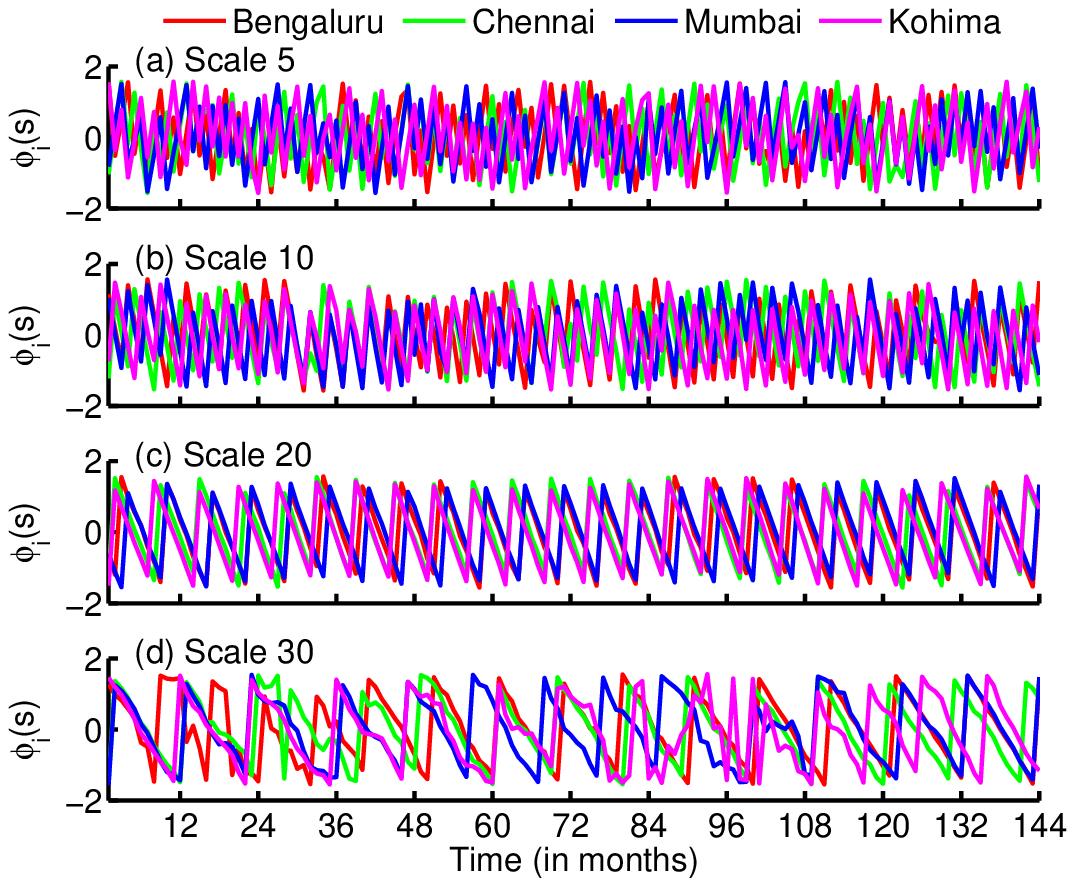}
\noindent \caption{\it The phase relation at different scales 
in Bengaluru, Chennai, Mumbai and Kohima.} 
\end{figure}

\newpage
\begin{figure}[h]
\includegraphics[width=1.0\textwidth]{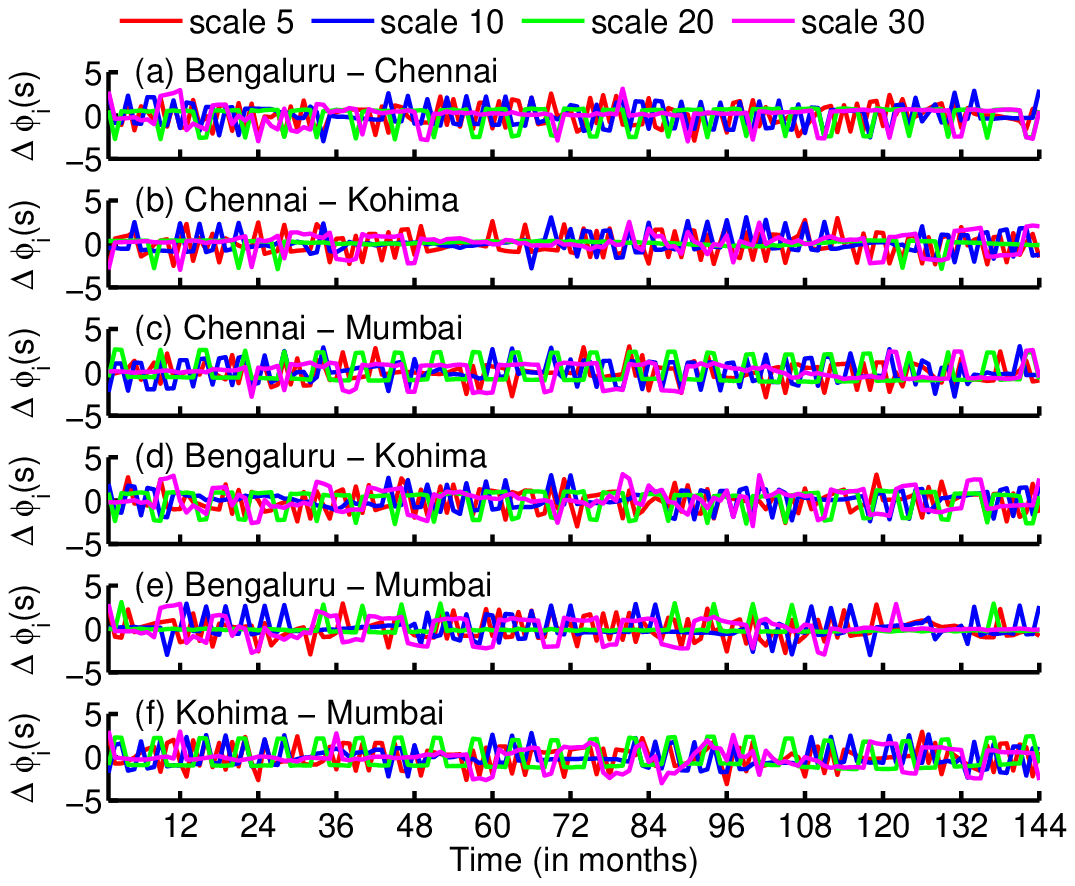}
\noindent \caption{\it The phase differences between the four major locations 
at different scales ranging from 5 to 30. 
The positive phase difference $\Delta \phi \equiv \phi_A -\phi_B > 0$ 
indicates that location $A$ is leading location $B$, while $\Delta \phi<0$ 
indicates that location $A$ is lagging behind location $B$.}
\end{figure}

\end{document}